\documentclass[useAMS,usenatbib]{mn2e}
\usepackage{graphicx}
 \usepackage{epstopdf}
 \usepackage{epsfig}
\usepackage{amsmath}

\newcommand{\mpcoh}{\,h^{-1}\,{\rm Mpc}}
\newcommand{\overbar}[1]{\mkern 1.5mu\overline{\mkern-1.5mu#1\mkern-1.5mu}\mkern 1.5mu}
\newcommand{\refedit}{}

\title{Using angular pair upweighting to improve 3D clustering measurements}
\author[Will J. Percival \& Davide Bianchi]{Will J. Percival$^{1}$\thanks{E-mail: will.percival@port.ac.uk}, Davide Bianchi$^{1}$\\
$^{1}$Institute of Cosmology and Gravitation,  University of Portsmouth, Dennis Sciama building, Portsmouth, PO1 3FX}
\begin{document}

\date{}

\pagerange{\pageref{firstpage}--\pageref{lastpage}} \pubyear{}

\maketitle

\label{firstpage}

\begin{abstract}
  Three dimensional galaxy clustering measurements provide a wealth of
  cosmological information. However, obtaining spectra of galaxies is
  expensive, and surveys often only measure redshifts for a subsample
  of a target galaxy population. Provided that the spectroscopic data
  is representative, we argue that angular pair upweighting should be
  used in these situations to improve the 3D clustering
  measurements. We present a toy model showing mathematically how such
  a weighting can improve measurements, and provide a practical
  example of its application using mocks created for the Baryon
  Oscillation Spectroscopic Survey (BOSS). Our analysis of mocks
  suggests that, if an angular clustering measurement is available
  over twice the area covered spectroscopically, weighting gives a
  $\sim$10--20\% reduction of the variance of the monopole correlation
  function on the BAO scale.
\end{abstract}


\section{Introduction}
\label{sec:intro}

The recent cosmological measurements of the Baryon Acoustic
Oscillation (BAO) at the percent level, and measurements of the
Redshift-Space Distortion (RSD) signals at the few percent level, in
the clustering of galaxies selected from the Baryon Oscillation
Spectroscopic Survey (BOSS; \citealt{Daw13}), part of the Sloan
Digital Sky Survey III \citep{Eis11}, have clearly demonstrated the
power of such measurements to understand the low-redshift Universe
\citep{alam16}. Over the next decade, surveys including eBOSS
\citep{Daw16}, DESI \citep{amir16a,amir16b}, and Euclid
\citep{laureijs11}, will provide an order of magnitude improvement on
these constraints, helping us to understand Dark Energy.

Given the significant investment in these spectroscopic surveys, it is
imperative to extract as much information as possible from them. One
avenue that has been relatively unexploited until now is the
possibility of using extended target sample data to enhance the
spectroscopic measurements of a subset of it. It is common to be in
the position of only having spectroscopic data for a subsample of
targets. For example, the VIMOS Public Extragalactic Redshift Survey
(VIPERS; \citealt{guzzo14}) spectroscopically observed $24$\,deg$^2$
of the $157$\,deg$^2$ Canada-France-Hawaii Telescope Legacy Survey
(CFHTLS) Wide photometric catalogue. Similarly, the BOSS DR9 analyses
presented in \citet{anderson12} used $3,275$\,deg$^2$ of spectroscopic
data, out of a target sample covering $9,274$\,deg$^2$. An obvious
question is ``how best to use the extra target data to enhance the 3D
clustering measurements?''. We argue here that angular pair
upweighting does this.

Angular pair upweighting has previously been used to partially correct
for spectroscopic completeness. It was used in \citet{hawkins03} to
correct for missing close pairs in the 2dFGRS \citep{colless03}. It
was also used in recent analyses of the VIPERS data: the pattern of
observations allowed by the VIMOS instrument meant that the VIPERS
galaxy clustering signal was strongly distorted by the spectroscopic
completeness. \citet{delaTorre13} and \citet{delaTorre16} used angular
pair weighted to correct for the incompletenesses, weighting pairs
counted in a 3D correlation function measurement by the ratio
$(1+w_p(\theta))/(1+w_s(\theta))$, where $w(\theta)$ is the angular
correlation function, subscript $s$ denotes the spectroscopic sample,
and $p$ the photometric sample.

We investigate the utility of angular pair upweighting, not to correct
for missing observations, but to improve on 3D clustering
measurements. In our companion paper \citep{bianchi17} we present a
new scheme to de-bias measurements from the effects of systematically
missing observations, showing that angular pair upweighting alone only
partially alleviates the effects. However, in \citet{bianchi17}, we
argue that if forms part of a more complicated scheme to correct for
missing observations that uses angular pair upweighting to recoup
lost signal-to-noise. The investigation described here shows that it
has the potential for wider application in the analysis of
spectroscopic galaxy surveys.

In Section~\ref{sec:unc-pairs}, we present a toy analytic model for
angular pair upweighting, for the simple situation of uncorrelated 3D
pair measurements in a single bin in $r$ and $\theta$. This is
extended to correlated pairs in Section~\ref{sec:corr-pairs}. In
Section~\ref{sec:mocks} we present an analysis of mock catalogues
designed to match the DR12 BOSS sample, showing how angular clustering
measurements from a wide area can enhance 3D clustering
measurements. Our results are discussed in
Section~\ref{sec:discussion}.

\section{Uncorrelated pair counts}  \label{sec:unc-pairs}

We first present an analytic toy model demonstrating the basic idea of
pair upweighting. We have matched the notation of \citet{ls93}, so we
can use their method for describing pair counts within surveys, and
our equations can be directly compared to theirs.

Consider $n_g$ objects angularly distributed within a region $\Omega$
with dimensionless geometric form factor $G_p(\theta)$, which encodes
the geometry of the survey (as in \citealt{ls93}) as a distribution of
pairs separated by $\theta\pm \Delta\theta/2$. Now consider a
representative sample of galaxies that have spectroscopic data,
denoted with a subscript $1$. For \refedit{Poisson} distributed pairs,
the expected number in an angular bin is given by,
\begin{equation}  \label{eq:Np1-noc}
  N_{p,1}\equiv\langle n_{p,1}\rangle=\frac{n_{g,1}(n_{g,1}-1)}{2}G_{p,1}(\theta).
\end{equation}
Given spectroscopic information, suppose that $m_{p,1}$ of these are
placed in the radial bin $r\pm \Delta r/2$ with selection probability
$G_r(r|\theta)$. For the selection probability, we drop the subscript
$1$ as we assume that this sample has representative selection. We
have from Bayes theorem that
\begin{equation}  \label{eq:bayes1}
  P(m_{p,1})=\int\,dn_{p,1}\, P(m_{p,1}|n_{p,1})P(n_{p,1}).
\end{equation} 
\refedit{For uncorrelated pair counts, $m_{p,1}\sim B(n_{p,1}, G_r(r|\theta))$
has a Binomial distribution. In order to avoid problems later if we
had zero pairs in an angular bin (we need at least one pair in each
angular bin, to be able to divide by the observed pair counts), we
assume that $n_{p,1}$ is drawn from a zero-truncated Poisson
distribution, rather than a Poisson distribution, so that}
\begin{equation}  \label{eq:zeropoiss_pdf}
  P(n_{p,1}=k|n_{p,1}>1)=\frac{N_{p,1}^k}{(e^{N_{p,1}}-1)k!},
\end{equation}
\refedit{and the expected number of pairs is slightly changed,}
\begin{equation}
  \langle n_{p,1}\rangle=\frac{N_{p,1}e^{N_{p,1}}}{e^{N_{p,1}}-1}.
\end{equation}
\refedit{Using a zero-truncated Poisson distribution rather than a
  Poisson distribution is simply a mathematical convenience, as we
  will only be concerned with the situation where $N_{p,1}$ is large.}
Using Eq.~(\ref{eq:bayes1}) to solve for the mean and variance of
$m_{p,1}$ gives that
\begin{equation}  \label{eq:Mp1-noc}
  \langle m_{p,1}\rangle=\frac{N_{p,1}G_r}{1-e^{-N_{p,1}}},
\end{equation}
and
\begin{equation}
  {\rm Var}(m_{p,1})=\frac{N_{p,1}G_r+N^2_{p,1}G^2_r}{1-e^{-N_{p,1}}}-\frac{N^2_{p,1}G^2_r}{(1-e^{-N_{p,1}})^2},
\end{equation}
where we have written $G_r=G_r(r|\theta)$ for simplicity. In the limit
of large $N_{p,1}$, this reduces to the standard Poisson result of
both the mean and variance equalling $N_{p,1}G_r$. Here, when
calculating the variance, we have ignored the effect of triplets, as
considered, for example, in \citet{ls93}, and assume that the dominant
error term is Poisson.

Now suppose that we introduce angular sample $2$ for which we do not
have radial information, but that has clustering that is statistically
identical to that of sample $1$. \refedit{For our toy model, we assume
  that $n_{p,2}$ is drawn from a zero-truncated Poisson distribution,
  that is independent of $n_{p,1}$ and has a probability density
  function as in Eq.~(\ref{eq:zeropoiss_pdf}), but with parameter
  $N_{p,2}$ rather than $N_{p,1}$.}  We will construct our estimator
by combining both samples, so that we now have
\begin{equation}  \label{eq:bayes2}
  P(\tilde{m}_{p,1})=\int\,dn_{p,1}\,dn_{p,2}\, 
    P(\tilde{m}_{p,1}|n_{p,1},n_{p,2}) P(n_{p,1}) P(n_{p,2}).
\end{equation} 
We wish to use the combined information from both samples to form a
maximum likelihood estimator for the expected number of angular pairs
in sample $1$. Provided that $G_{p,1}(\theta)$ and $G_{p,2}(\theta)$
have similar distributions, then we can construct this by simply
summing the regions together
\begin{equation}
  \tilde{n}_{p,1}\equiv \frac{A_1(n_{p,1}+n_{p,2})}{A_1+A_2},
\end{equation}
where $A_1$ and $A_2$ are the relative areas covered by samples $1$
and $2$ respectively. Angular pair upweighting treats
$\tilde{n}_{p,1}/n_{p,1}$ as a weight to be applied to $m_{p,1}$ to
give $\tilde{m}_{p,1}$
\begin{equation}
  \tilde{m}_{p,1}=\frac{A_1(n_{p,1}+n_{p,2})}{n_{p,1}(A_1+A_2)}m_{p,1},
\end{equation}
and we see that the mean is unchanged. i.e.
\begin{equation}
  \langle \tilde{m}_{p,1}\rangle=\frac{N_{p,1}G_r}{1-e^{-N_{p,1}}}.
\end{equation}
Working in the limit where $N_{p,1}$ and $N_{p,2}$ are large so that
$\langle 1/N_{p,1} \rangle\simeq 1/\langle N_{p,1} \rangle$,
$e^{-N_{p,1}}\to0$ and $e^{-N_{p,2}}\to0$, the variance is changed to
\begin{displaymath}
   {\rm Var}(\tilde{m}_{p,1})=\frac{A_1^2G_r}{(A_1+A_2)^2}
     \Bigg[N_{p,1}+N_{p,2}G_r+
\end{displaymath}
\begin{equation}
    \left(2N_{p,2}+\frac{N_{p,2}}{N_{p,1}}+\frac{N^2_{p,2}}{N_{p,1}}\right)(1-G_r)\Bigg].
\end{equation}
Taking the limit of large $N_{p,1}$, and rewriting areas in terms of
$N_{p,1}$ and $N_{p,2}$, the difference between this and the original
variance reduces to the simple form
\begin{equation}  \label{eq:Vimprov}
  \frac{N_{p,1}G_r-{\rm Var}(\tilde{m}_{p,1})}{N_{p,1}G_r}=\frac{G_r}{1+N_{p,1}/N_{p,2}},
\end{equation}
and we clearly see that
${\rm Var}(\tilde{m}_{p,1})<{\rm Var}(m_{p,1})$ for all
$N_{p,2}>0$. The numerator of $G_r$ in Eq.~(\ref{eq:Vimprov}) shows
that the improvement in the variance is proportional to the fraction
of pairs entering each radial bin.  The denominator shows that the
\refedit{limiting} maximum fractional improvement, \refedit{which
  equals} $G_r$, is logarithmically reached in the limit of large
$N_{p,2}/N_{p,1}$.

\section{Correlated pair counts}  \label{sec:corr-pairs}

The relation between the 3D correlation function and its angular and
radial components can be written
\begin{equation}
  1+\xi(r,\theta)=(1+\xi(r|\theta))(1+w(\theta)),
\end{equation}
where $w(\theta)$ is the angular correlation function, and
$\xi(r|\theta)$ is the radial correlation function for pairs within
this angular bin. As explained in \citet{ls93}, because our sample
geometry limits the distribution of pairs in both angular and radial
directions, the relation between the measured clustering and the true
correlation function requires an ``Integral Constraint''. Thus, with
clustering, Eq.~\ref{eq:Np1-noc} becomes
\begin{equation}  \label{eq:Np1-wc} 
  N_{p,1}=\frac{n_{g,1}(n_{g,1}-1)}{2}G_{p,1}(\theta)\frac{1+w(\theta)}{1+w_\Omega},
\end{equation}
where
\begin{equation} 
  1+w_\Omega=\int_\Omega\,d\Omega\, G_{p,1}(\theta)(1+w(\theta)).
\end{equation}
Similarly in the radial direction, Eq.~(\ref{eq:Mp1-noc}) becomes
\begin{equation}
  \langle m_{p,1}\rangle=\frac{N_{p,1}G_r(r|\theta)(1+\xi(r|\theta))}
    {(1+\xi_r)(1-e^{-N_{p,1}})},
\end{equation}
where
\begin{equation}
  1+\xi_r=\int_r\, dr\, G_r(r|\theta)(1+\xi(r|\theta).
\end{equation}

\begin{figure} 
\includegraphics[width=85mm]{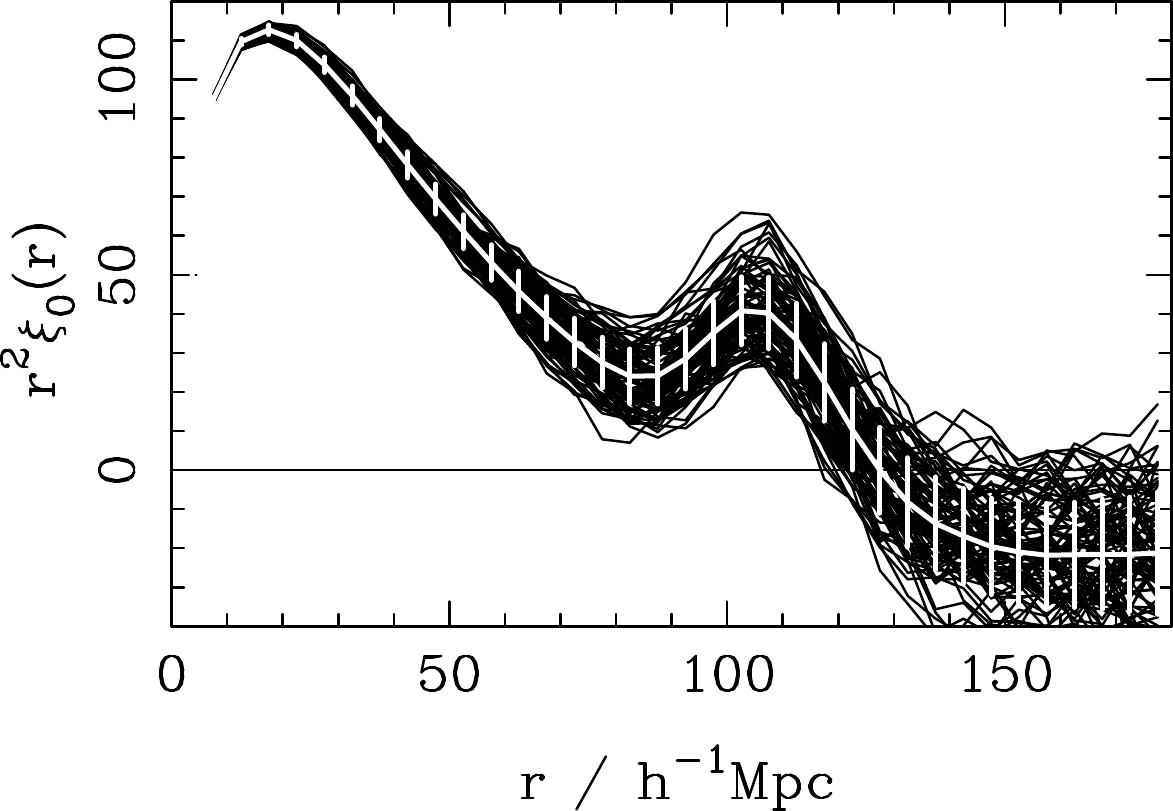} 
\caption{Monopole moments of the correlation function rebinned into
  bins of width $\Delta r=5.5\mpcoh$, calculated from 100 of the 1000
  QPM mocks used, matching the NGP component of the DR12 BOSS CMASS
  sample (black lines). The overplotted white line shows the mean,
  with 1$\sigma$ errors. \label{fig:cmass-xi0}}
\end{figure}

Thus we see that the effect of clustering, including both the
correlation functions and integral constraints, can be absorbed into
the geometric factors. Defining
\begin{eqnarray}  \label{eq:improv}
  G'_{p,1}&\equiv&G_{p,1}(\theta)\frac{1+w(\theta)}{1+w_\Omega},\\
  G'_r&\equiv&G_r(r|\theta)\frac{1+\xi(r|\theta)}{1+\xi_r},
\end{eqnarray}
we can use the results of Section~\ref{sec:unc-pairs}, with the
transform $G_{p,1}\to G'_{p,1}$, and $G_r\to G'_r$.
 
\section{Analysis of BOSS DR12 mocks}  \label{sec:mocks}

We have tested the utility of angular pair upweighting using the mock
catalogues created to match the Data Release 12 (DR12; \citealt{DR12})
galaxy sample of BOSS.\footnote{The choice of using these catalogues
  was driven by the desire to have a large number of public mock
  galaxy catalogues from which we can calculate variances, rather than
  any desire to test this specific sample.}  Specifically we have
analysed 1000 mock catalogues created from Quick Particle Mesh (QPM;
\citealt{white14}) simulations designed to mimic the DR12 CMASS galaxy
sample. We also analyse 1000 mock catalogues created using the
MultiDark-Patchy (hereafter MD-Patchy; \citealt{kitaura16}) technique
designed to mimic the DR12 LOWZ galaxy sample. See \citet{reid16} for
details of the BOSS galaxy samples. MD-Patchy uses second-order
Lagrangian perturbation theory and a stochastic halo biasing scheme
calibrated using the Multi-Dark simulation, while QPM selects haloes
from low-resolution particle mesh simulations such that they match the
expected 1-point and 2-point statistics. For both methods, halos are
populated using a Halo Occupation Distribution (HOD) model to
construct galaxy density fields. Each mock is then sampled to match
both the angular and radial selection functions of the survey. Thus we
compare results using two mock catalogue production schemes and two
different target samples. As we only wish to provide a
proof-of-concept, we do not perform reconstruction, a technique to
remove the smearing effect of bulk motions - combining angular pair
upweighting with reconstruction is discussed further in
Section~\ref{sec:discussion}. We only consider the North Galactic Cap
(NGC) part of the survey: the CMASS sample contains $\sim570k$
galaxies, while the LOWZ sample contains $\sim250k$ galaxies.

\begin{figure} 
\includegraphics[width=85mm]{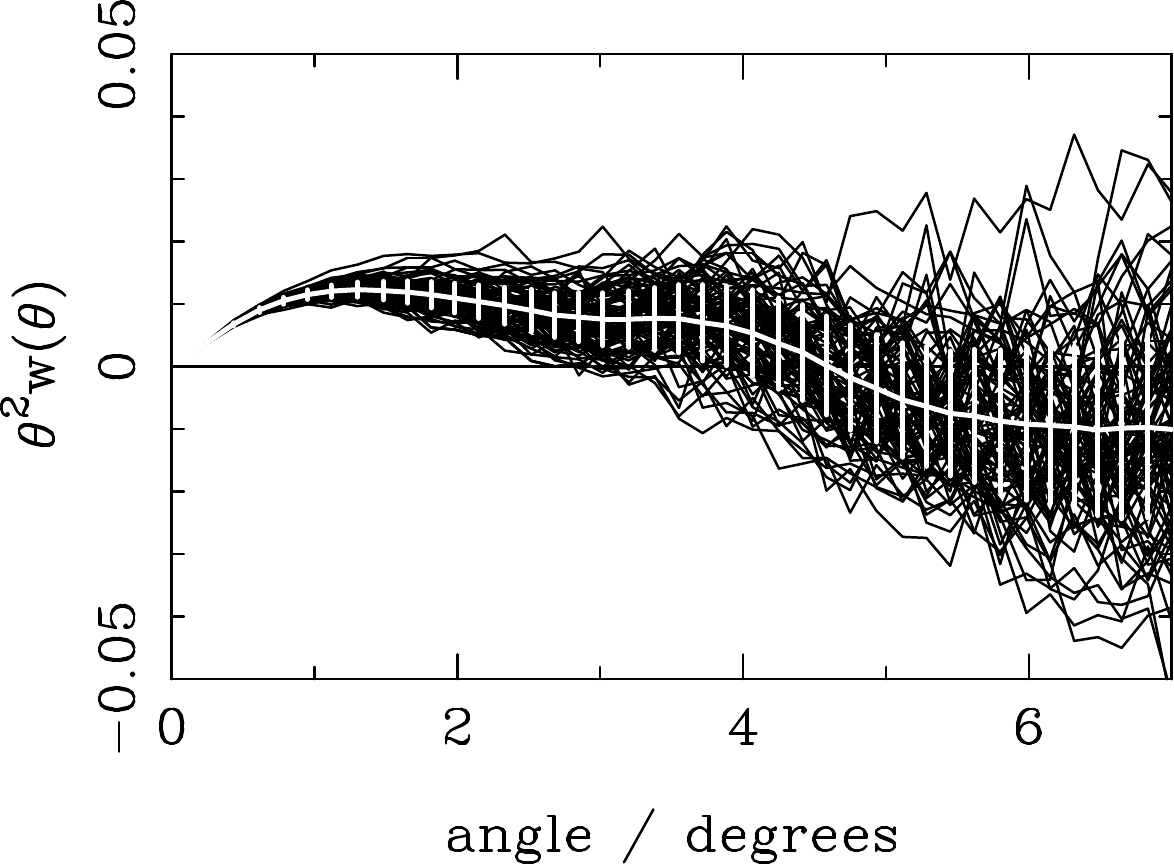} 
\caption{Lines as for Fig.~\ref{fig:cmass-xi0}, but now for the
  angular correlation function of 100 of the QPM CMASS mocks, binned
  into bins of width $\Delta \theta=0.16$\,degrees. 
  \label{fig:cmass-wt}}
\end{figure}

To quantify the survey masks of both the CMASS and LOWZ samples, we
used a catalogue of points chosen randomly to sample the survey
regions, containing 20$\times$ as many points as the number of
galaxies in each sample. Using the galaxy and random catalogues we
have measured galaxy-galaxy (DD), galaxy-random (DR) and random-random
(RR) pair-counts. The pair-counts were binned into $180\times270$ bins
with separation $\Delta r = 1\mpcoh$, $0<r<180\mpcoh$ and linearly
spaced in $0<\theta<9$\,degrees for CMASS and $0<\theta<21$\,degrees
for LOWZ. We include pairs that are within the angular range
considered, but have $r>180\mpcoh$ in the final bin in $r$, so that
our angular clustering can be measured as if there were no radial
information. The range of angles considered is driven by calculating
$w(\theta)$ for the pairs with the separations of interest, and so is
different for LOWZ and CMASS samples. We have tried a number of
resolutions for the angular pair count bins, and find that it does not
significantly influence the results: for simplicity we downsample the
angular pair counts by a factor of 5 for the results presented. As we
only consider the monopole moment, we do not additionally bin in
$\mu$, the cosine of the angle to the line-of-sight, although we would
expect angular pair weighting to help improve the full 3D correlation
function for all $\mu$ except $\mu=1$, which corresponds to $\theta=0$
and where the angular correlations are formally both zero. Using the
\citet{ls93} estimator, we used these pair counts to measure the
monopole moment of the correlation function for each mock
catalogue. For a selection of the QPM CMASS mocks, these are plotted
against their mean and variance in Fig.~\ref{fig:cmass-xi0}.

We wish to contrast the measured variance of the monopole correlation
function with that obtained using angular pair upweighting as if we
had a wider area of projected angular data. We have calculated the
angular correlation function for mock $i$, $w_i(\theta)$, where we
take the binned pair counts described above for $DD_i(\theta)$,
$DR_i(\theta)$ and $RR_i(\theta)$ split by $\theta$ and use the
\citet{ls93} estimator: the angular correlation functions for a
selection of the QPM CMASS mocks are plotted in
Fig.~\ref{fig:cmass-wt}. In order to consider an angular clustering
measurement made as if from a wider survey, we average angular
pair-counts from the mock of interest and $N_{\rm proj}-1$
additional mocks, creating a lower-variance estimate of the $DD$ and
$DR$ angular pair-counts:
\begin{equation}
  \overbar{DD}_i(\theta)=\frac{1}{N_{\rm proj}}\sum_i^{i+N_{\rm proj}}DD_i(\theta),
\end{equation}
and similarly for $\overbar{DR}_i$. We determine the ratios of counts in
the low variance estimate to that from the individual mock,
$\overbar{DD}_i(\theta)/DD_i(\theta)$ and $\overbar{DR}_i(\theta)/DR_i(\theta)$ in
bins of width $\Delta \theta=0.16$\,degrees. We use this ratio to
weight the paircounts for each bin in $r$ and $\theta$, so our
weighted estimate for $DD_i(r)$ is given by
\begin{equation}
  \overbar{DD}_i(r) = \sum_{\theta} DD_i(r,\theta) \frac{\overbar{DD}_i(\theta)}{DD_i(\theta)}.
\end{equation}
The monopole moment of the correlation function was then recalculated
for each mock using $\overbar{DD}(r)$ and $\overbar{DR}(r)$. $RR(r)$ is
unchanged as we use the same random catalogue for all mocks. The
average weighted monopole matches the unweighted average with
differences at the level expected from noise.

\begin{figure} 
\includegraphics[width=85mm]{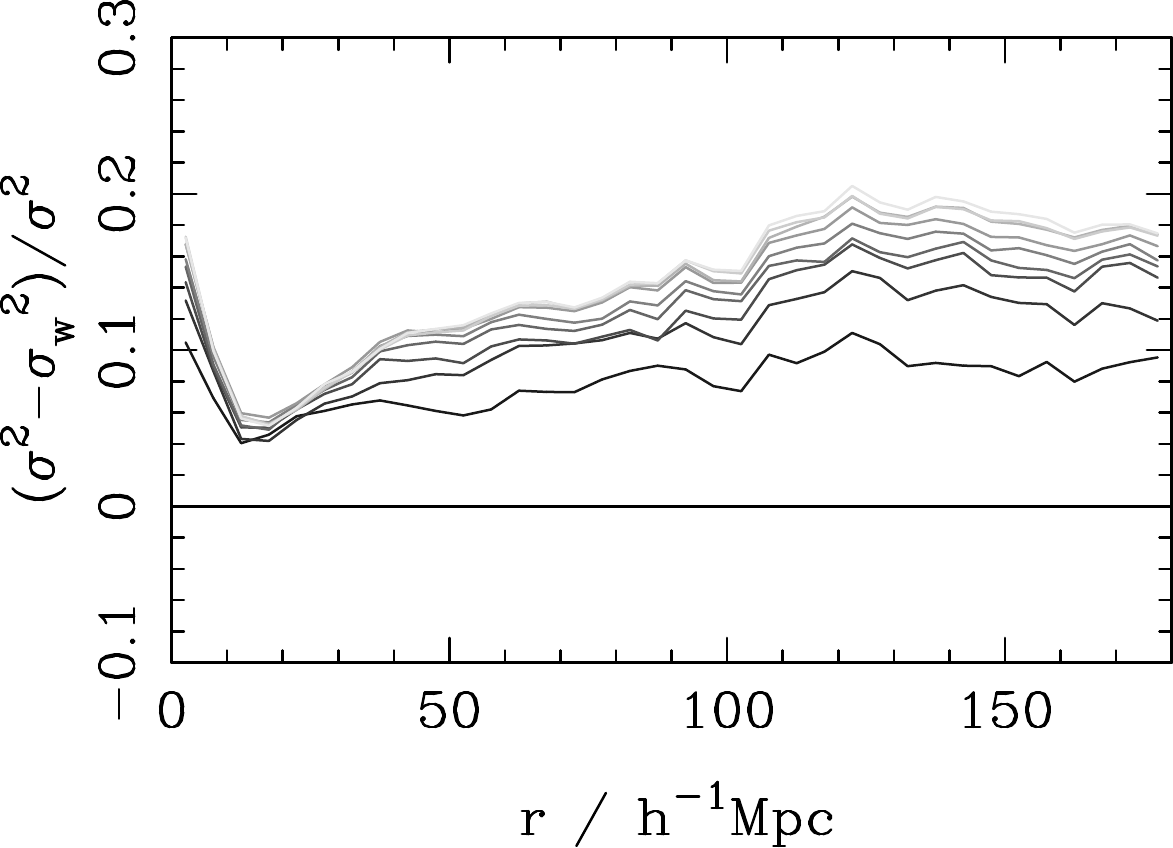} 
\caption{The fractional improvement in the variance for the CMASS DR12
  NGP mocks assuming there is additional information on the angular
  correlation function corresponding to $N_{\rm proj}=2...10$ -
  i.e. between twice and 10$\times$ the area. $\sigma^2$ is the
  unweighted variance, and $\sigma_w^2$ is the variance after
  angular pair upweighting. \label{fig:cmass-variance}}
\end{figure}

\begin{figure} 
\includegraphics[width=85mm]{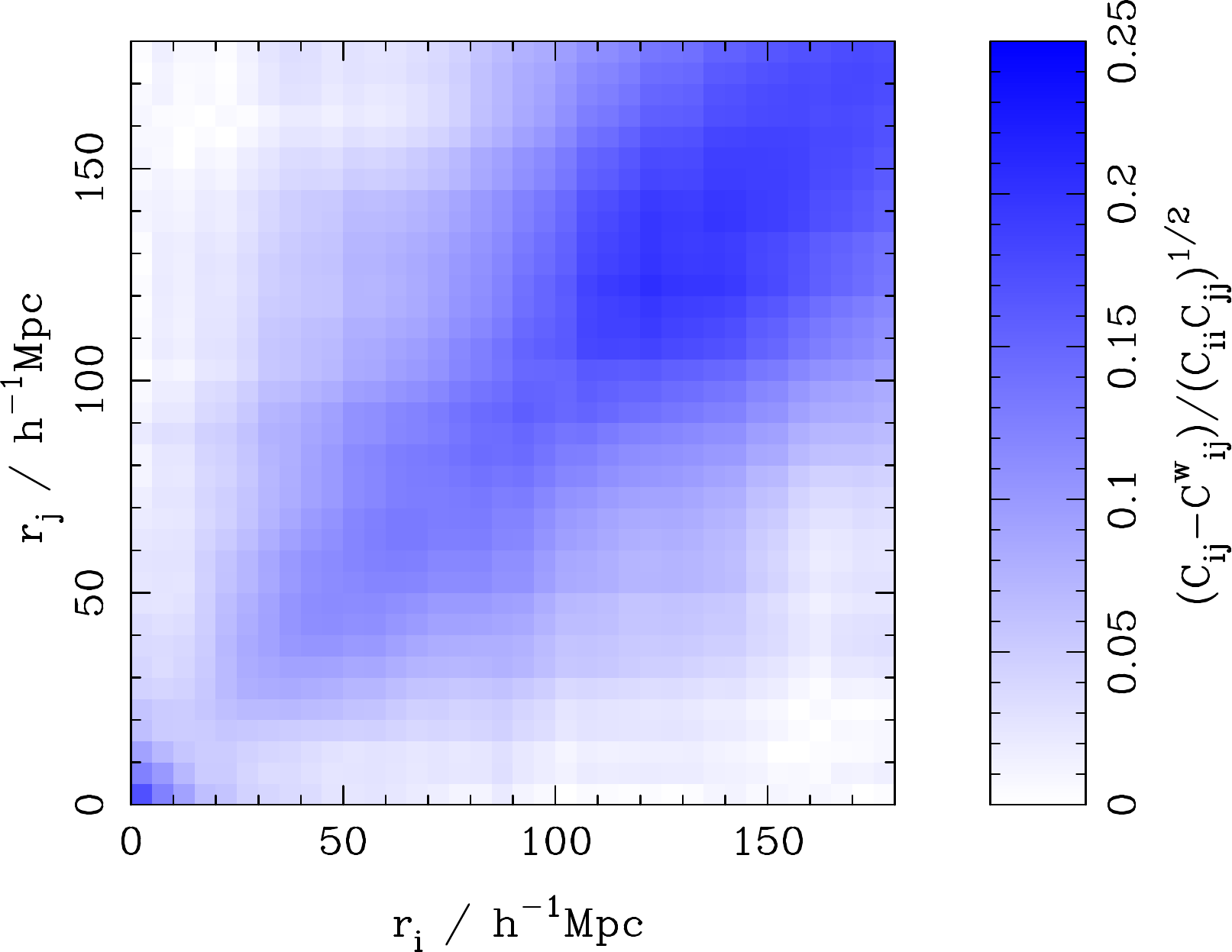} 
\caption{The fractional improvement in the correlation matrix for the CMASS DR12
  NGP mocks assuming there is additional information on the angular
  correlation function corresponding to 10$\times$ the area. $C_{ij}$
  is the element of the covariance matrix for $\xi_0$ between bins $i$
  and $j$, and $C^w_{ij}$ is the covariance matrix element for the
  weighted correlation function. \label{fig:cmass-mat}}
\end{figure}

Fig.~\ref{fig:cmass-variance} presents the fractional improvement in
the variance obtained for the CMASS DR12 QPM mocks. We show the
potential improvement as if we had between $2\times$ and $10\times$
the angular data with which to weight the 3D pair counts. Although
unrealistic for BOSS, as the area covered by the NGP sample is already
$\sim7,000$\,deg$^2$, this demonstrates the logarithmic improvement to
the 3D clustering measurements possible with increasing amounts of
angular data, matching the asymptotic behaviour of
Eq.~(\ref{eq:improv}). Fig.~\ref{fig:cmass-mat} shows that the
improvement in the covariance extends to the off-diagonal
elements. For simplicity, we only show the improvement for the case of
having $10\times$ the angular data with which to weight the 3D pair
counts.

\begin{figure} 
\includegraphics[width=85mm]{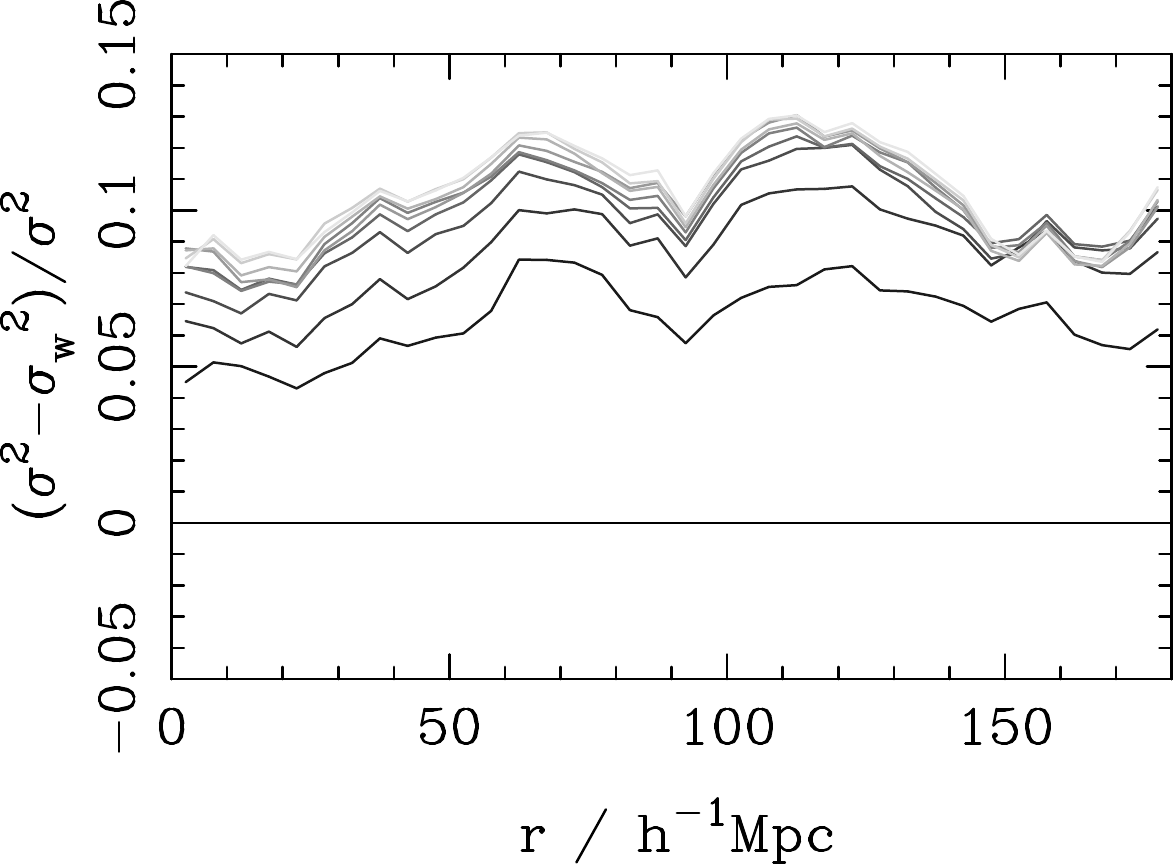} 
\caption{As Fig.~\ref{fig:cmass-variance}, but now for the MD-Patchy
  LOWZ mocks. \label{fig:lowz-variance}}
\end{figure}
The fractional improvement for the LOWZ DR12 MD-Patchy mocks is shown
in Fig.~\ref{fig:lowz-variance}. As in Fig.~\ref{fig:cmass-variance},
we show the potential improvement as if we had between $2\times$ and
$10\times$ the angular data with which to weight the 3D pair
counts. The improvement for LOWZ is slightly worse than that for
CMASS, asymptoting to a $\sim$10\% improvement on BAO scales for large
angular target areas.

\section{Discussion}  \label{sec:discussion}

We have presented analytic and numerical arguments that angular-pair
upweighting is an efficient method to enhance the 3D clustering
measurements from spectroscopic surveys, when angular clustering data
is available over a larger area than that observed
spectroscopically. Results from both the analytic derivation and from
mock catalogues show improvement in the error on the resulting 3D
correlation function measurements. We have also shown that angular
pair upweighting gives rise to unbiased clustering measurements.

In essence, the method works by exploiting the correlation between
angular and 3D clustering: by dividing the spectroscopic pair counts
by the angular pair counts, we remove this component of the noise, and
we then multiply by an estimator with the same mean but lower noise to
get back to an unbiased estimate of the 3D pair counts. In terms of
radial and angular modes, we are using additional angular data to
reduce the error on the contribution of these modes to the 3D
clustering measurement.

Our analysis of mock catalogues shows a larger improvement than that
in the toy analytic model, \refedit{and the level of improvement is
  different for the LOWZ and CMASS samples}: we expect this is because
the monopole correlation function measurements are correlated for the
mock data across different $r$, unlike the toy model. Angular
projection acts as a smoothing of the monopole with an
asymmetric kernel, and so the correlations between angular and 3D
$\xi$ will be increased by coherence in the 3D $\xi$- i.e. the
correlations in the clustering signal mimic slightly those of the
angular projection. \refedit{There will also be differences caused by
  changes in the relative importance of angular compared to radial
  clustering in the final measurement, due to survey geometry and
  redshift-space effects.}

There are some caveats to the analysis presented: in particular, we
have only tested the efficiency of pair upweighting using binned
data. It would be possible to provide a weight for every galaxy pair
using the actual separation (this was the procedure adopted in
\citealt{bianchi17}). However, using the binned counts is conservative
and, given the number and size of the BOSS mocks used, using
individual weights for each pair would have been prohibitively
expensive.

It is not obvious how the proposed upweighting would affect clustering
measurements post-reconstruction of the density field, which has
become an essential part of BAO analyses. However, it should be
possible to reap the benefits of pair upweighting, while including the
information from reconstruction by following the recent work of
\citet{sanchez17}. This suggested that the combination of
post-reconstruction and weighted pre-recon measurement could be
performed after parameter measurement, while retaining the information
from both. Practically, one would calculate upweighted
pre-reconstruction clustering and non-weighted post-reconstruction
clustering, and combine the measurements of cosmological parameters
from both to form combined parameter measurements, allowing for
correlations between measurements. The angular upweighting would
reduce the correlation between pre- and post-reconstruction
measurements, enhancing the combined signal.

For obvious reasons, the method works best where there is a strong
correlation between 3D and angular clustering measurements, so it will
provide a more significant improvement for thinner redshift
shells. Thus, if photometric redshifts were provided for all target
galaxies in the sample, including those with spectroscopic redshifts,
these could be used to split the sample into thinner redshift shells
before applying the angular upweighting. It would be possible to work
with shells in photo-z, or using an interpolation scheme across
redshift. This would enhance the efficiency of the weights.

\refedit{As constructed, the method will work where the samples used
  are of the same underlying objects. Enhancing the 3D clustering
  using angular samples of a different population would be possible,
  but would come at the expense of having to model the
  cross-correlation. The method will help to improve both shot-noise
  and sample variance errors: for a narrow spectroscopic survey,
  upweighting using a wider angular sample of target galaxies will
  improve the sample variance error, while upweighting a sparse
  spectroscopic sample of a wide survey using a denser sample of
  angular pair counts over the same area would improve the shot
  noise.}

\section*{Acknowledgments}

WJP and DB acknowledge support from the European Research Council
through the Darksurvey grant 614030. WJP also acknowledges support
from the UK Science and Technology Facilities Council grant
ST/N000668/1 and the UK Space Agency grant ST/N00180X/1. We thank
Ashley Ross and Hector Gil-Marin for useful conversations. Numerical
computations were done on the Sciama High Performance Compute (HPC)
cluster which is supported by the ICG and the University of
Portsmouth.

\label{lastpage}

\begin{thebibliography}{99}

\bibitem[{{Alam et al.}(2015)}]{DR12}
  Alam S., et al., 2015, ApJS 219, 12

\bibitem[{{Alam et al.}(2016)}]{alam16}
  Alam S., et al., 2016, arXiv:1607.03155

\bibitem[{{Amir et al.}(2016a)}]{amir16a}
  Amir S., et al., 2016a, arXiv:1611.00036

\bibitem[{{Amir et al.}(2016b)}]{amir16b}
  Amir S., et al., 2016b, arXiv:1611.00037

\bibitem[{{Anderson et al.}(2012)}]{anderson12}
  Anderson L., et al., 2012, MNRAS 427, 3435

\bibitem[{{Bianchi \& Percival}(2017)}]{bianchi17}
  Bianchi D., Percival W.J., 2017, arXiv:1703.02070

\bibitem[{{Colless et al.}(2003)}]{colless03}
  Colless M., et al., 2003, [astro-ph/0306581]

\bibitem[{{Dawson et al.}(2013)}]{Daw13}
  Dawson K., et al., 2013, AJ, 145, 10

\bibitem[{{Dawson et al.}(2016)}]{Daw16}
  Dawson K., et al., 2016, AJ, 151, 44

\bibitem[{{de la Torre et al.}(2013)}]{delaTorre13}
  de la Torre S., et al., 2013, A\&A, 557, A54

\bibitem[{{de la Torre et al.}(2016)}]{delaTorre16}
  de la Torre S., et al., 2016, arXiv:1612.05647

\bibitem[{{Eisenstein et al.}(2011)}]{Eis11}
  Eisenstein D.J., et al., 2011, AJ, 142, 72

\bibitem[{{Guzzo et al.}(2014)}]{guzzo14}
  Guzzo L., et al., 2014, A\&A, 566, 108

\bibitem[{{Hawkins et al.}(2003)}]{hawkins03}
  Hawkins E., et al., 2003, MNRAS 346, 78

\bibitem[{{Kitaura et al.}(2016)}]{kitaura16}
 Kitaura F.-S., et al., 2016, MNRAS, 456, 4156

\bibitem[{{Landy \& Szalay}(1993)}]{ls93}
  Landy S.D., Szalay A.S., 1993, ApJ 412, 64

\bibitem[{{Laureijs}(2011)}]{laureijs11}
  Laureijs R., et al., 2011, arXiv:1110.3193

\bibitem[{{Reid et al.}(2016)}]{reid16}
 Reid B., et al., 2016, MNRAS, 455, 1553

\bibitem[{{Sanchez et al.}(2016)}]{sanchez17}
  Sanchez A., et al., 2017, MNRAS 464, 1493

\bibitem[{{White et al.}(2014)}]{white14} 
  White M., Tinker J. L., McBride C. K., 2014, MNRAS, 437, 2594


\bibliographystyle{mnras}
\end{thebibliography}
\end{document}